# ESTABLISHING DATA WAREHOUSE TO IMPROVE STANDARDIZE HEALTH CARE DELIVERY: A PROTOCOL DEVELOPMENT IN JAKARTA CITY

Verry Adrian, Jakarta Department of Health, dr.verry@gmail.com

Intan Rachmita Sari, Jakarta Department of Health, intanrachmita@gmail.com

Hardya Gustada Hikmahrachim, Jakarta Department of Health, hardyagustada@yahoo.com

**Abstract:** Jakarta is a metropolitan city and among the most dense city in Indonesia. Jakarta has 12 major indicators of standardize health care delivery (*Standard Pelayanan Minimum* or SPM) derivates from Ministry of Health consists of services related to maternal and neonatal health, school-aged population, working-age population, elderly population, some specific conditions (hypertension, diabetes, tuberculosis, HIV), and also mental health. We planned to construct a data warehouse to provide a single integrated data center. In the first phase (2021), we improve the system by giving responsibility to the health Sub Department of Health of Administrative City for direct data input into a data warehouse. This period also let an introduction and adaptation to new data collection schemes by using a single entry for the first time. The basic platform use for this system is District Health Information System 2 (DHIS-2), an open-source platform that has been used worldwide, including Ministry of Health Republic of Indonesia. The major advantage of this data warehouse is the simplicity and convenience to collect a wide data from a different source and presenting it faster than using the conventional system. Less data contradiction was also found between health programs with intersecting data. During this transition phase, a double-work is made as data should be input to both the DHIS-2 system by Jakarta and the National Ministry of Health system, but an integration process is ongoing, and hopefully that in 2022 single data entry can be established.

## Introduction

Jakarta is a metropolitan city and also acts as the capital city of the Republic of Indonesia. The population of Jakarta City in 2020 is around 10.5 million occupying an area of 699.5 km square. Various efforts have been made to improve the health status in Jakarta. Following the instruction from The Central Government, there were twelve major indicators of standardize health care delivery, named *Standard Pelayanan Minimum* (SPM). SPM consists of services related to maternal and neonatal health, school- aged population, working-age population, elderly population, some specific conditions (hypertension, diabetes, tuberculosis, HIV), and also mental health. SPM are government affairs that must be pursued

Since the beginning of the regulation on SPM, there have been several problems related to recording and reporting of SPM data. Before the SPM Data Warehouse was created, data reporting was done manually from district health office to provincial health office via excel sent via email or google spread sheet, as well as the verification and validation processes were also done manually. This prolonged the data collection process, and made it difficult for manager program to recapitulate data. Including the data visualization process which also complicates the program, so that manager of program must spend more effort to analyze existing SPM data.

Furthermore, there are several data elements for each indicator that must be verified by two different programs, such as data on immunization for children under five in the indicator for children who receive health services. So far, the immunization program and the child health program in the health office have always reported basic immunization data for toddlers with different data values, so that outsiders always question the quality of the data in the end. With the SPM Data Warehouse, we can adjust the authority of each existing data element, so that there is no longer the publication of different data values with the same data element.

Accurate and actual data should be gathered to support health program implementation. On the other hand, data collection in a low-resource situation is another challenge in Jakarta. It impossible to draw health problems, and further allocate health resources, if no reliable data available.[2] Quality data will help the government to formulate objective and accurate policies to shoot at existing problems.

In the middle of 2020, we start to digitize health data. We planned to construct a data warehouse to provide a single integrated data center. The initial purpose is to optimize data collection for SPM-related programs and further use in all health-related data. Jakarta might be the pioneer in applying big data to public health in a practice at province level in Indonesia. The development of this data warehouse is under monitoring of Ministry of Health and in collaboration with the University of Oslo. This paper was written to share some strategies and experience in preparing concepts of data warehouse, thus might give insights to other cities in developing countries.

### Goal

The goal in building public health data warehouse in Jakarta city are:

1. Speed up the process of reporting and collecting SPM data

2. Make it easy for data visualization and analysis so that presenting comprehensive health data as a basis for health policy making in DKI Jakarta

3. Integration and synchronization of existing health data in DKI Jakarta

### Profile of Jakarta City

In designing public health data collection, demographic assessment is an important step. Jakarta is a province – known as a special capital region – consist of five administrative cities (central, north, south, east, and west Jakarta) and an administrative regency (Seribu Island), but popularly known as Jakarta City. It is the smallest province but the densest population in Indonesia, approximately 16.262 per square kilometer. In 2019, its gross domestic product (GDP) in total is $660 million or $55 per capita. Its Human Development Index (HDI) is 0.807, classified as very high (2019).

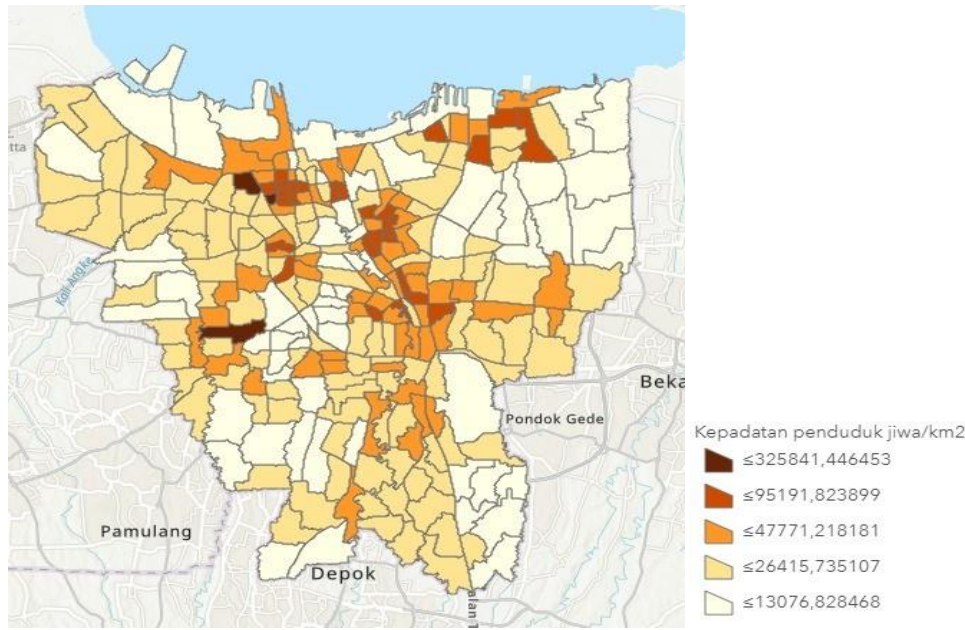

**Figure 1. Jakarta Map and Its Population Density**

A lot of things need to be done to improve Jakarta health status. In 2019, both maternal and neonatal mortality rates are still high (55.4 per 100.000 livebirth and 5.48 per 1000 livebirth, respectively). Services for noncommunicable diseases, such as hypertension, diabetes, and mental health also need improvement. Besides, currently, Jakarta is still in the fight for COVID-19 pandemics.

**The current system for data collection**

The major sources of health data in Jakarta are health program data and healthcare services data. The majority of data flow is in a conventional scheme by using e-mail or online documents which can be accessed by limited persons. Data were recapitulated and visualized manually by the data manager. Several programs that are involved in one data element may report different aggregate data. The integration of data is the responsibility of the Data and Information Section of Jakarta Department of Health. Following an urgency to develop a system for data utilization, at 2020 Jakarta designing a framework for public health data integration as seen in Figure 2.

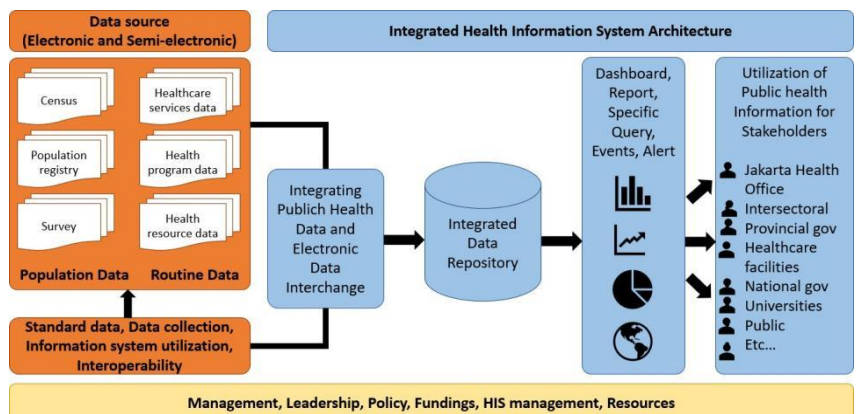

**Figure 2. Public Health Data Integration Framework Developed for Jakarta**

The focus on the first phase of development is on the box "Integrating Public Health Data and Electronic Data Interchange" from the current data source and then develop an integrated data repository. Data visualization in different forms such as dashboard, report, specific query, events, and alert are an innovation to improve health program evaluation and monitoring for stakeholders. This framework is the main goal for the next 5 years.

### Conceptual Framework

In establishing a system for building a data warehouse for data collection, a conceptual framework is the main foundation. We modified a framework from Public Health Surveillance and Action (McNabb et al, 2002)[3] as seen below.

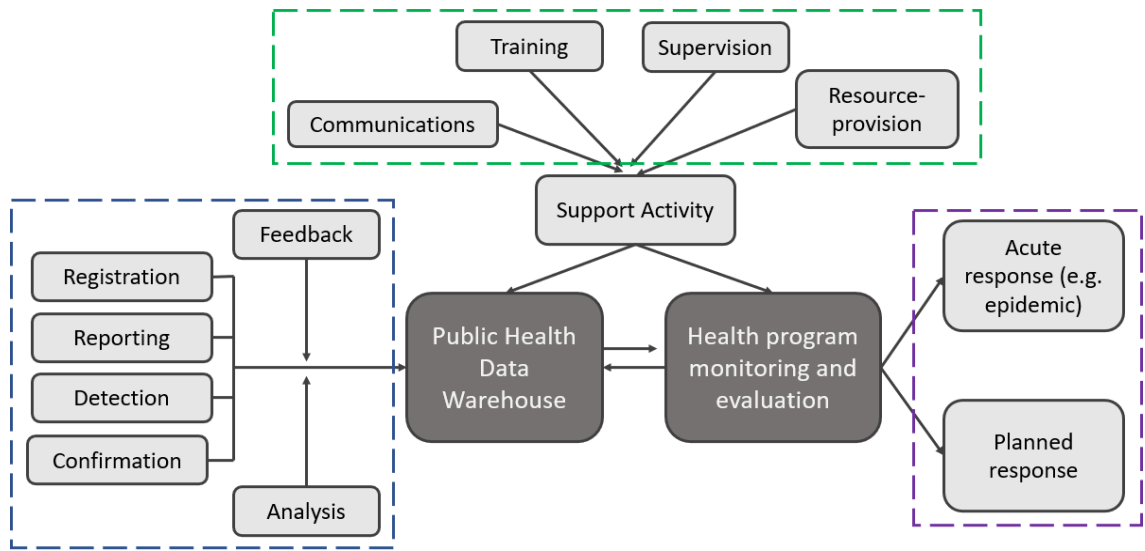

**Figure 3. Data warehouse development framework. Adapted from McNabb, et al (2002)**

Based on the framework, the input for the data warehouse (blue dash line box) is the responsibility of Puskesmas, covering district and sub-district areas. Data warehouse system provides the ability to register, report, detect and confirm data with direct analysis dan feedback. In this section, the data source should be of a high-quality. On the support activity section (green dash line box), Sub Department of Health of Administrative City (sudinkes) acts to ensure that the process is always in line with the framework by doing tight supervision and close communication. Training and resource-provision might be conducted in collaboration with Jakarta Department of Health (dinkes).

Data collected in a data warehouse are utilized for rapid health program monitoring and finally the main outcome: giving response to both acute situation (such as epidemic) or a planned response. In the first years of this data warehouse development, we are focusing to utilize for a planned response, according to 12 indicators of SPM.

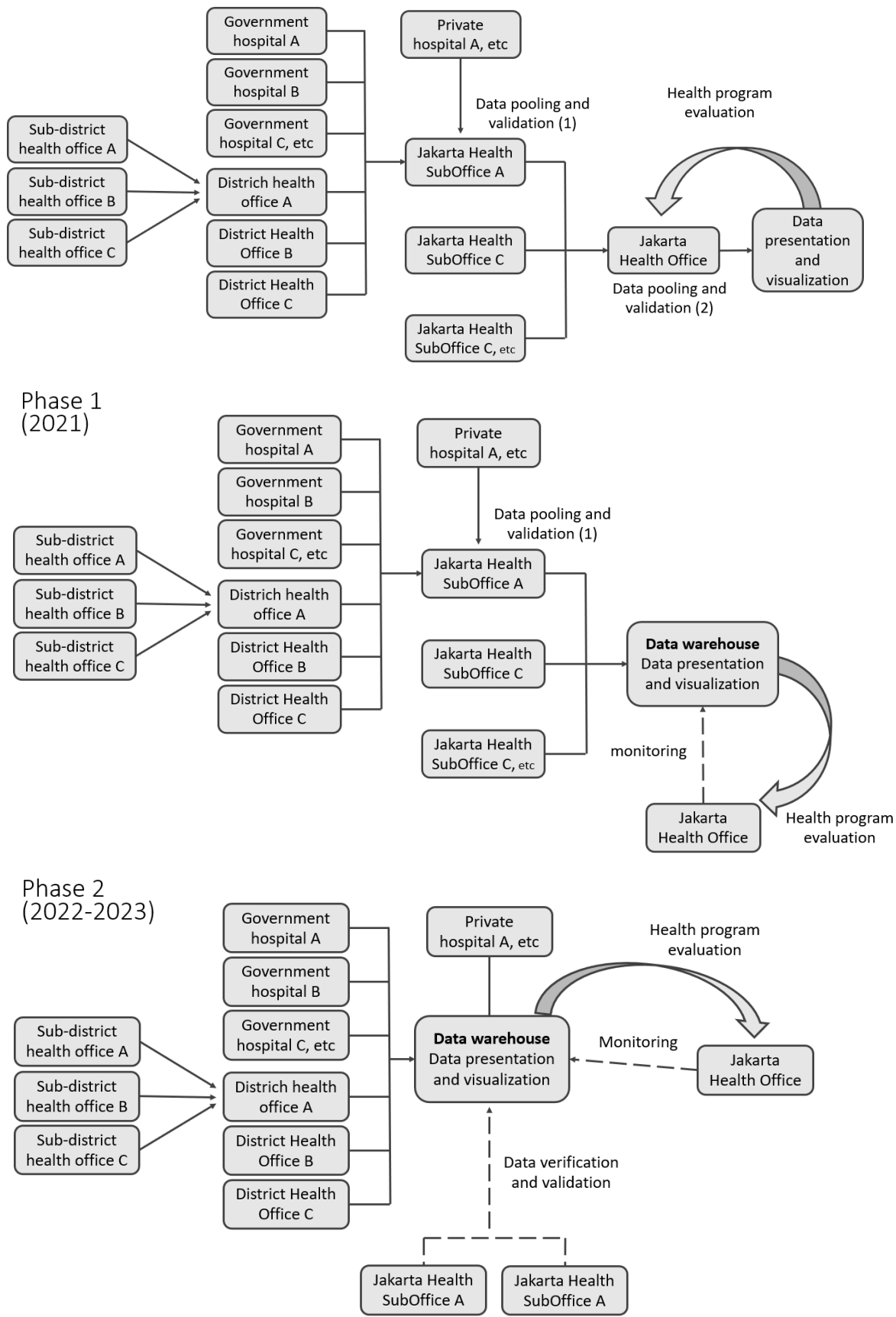

**Figure 4. Data collection and management in Jakarta. A) current situation, B) Design for Phase 1, C) Design for Phase 2. Note: Jakarta Health Office = Jakarta Department of Health; Jakarta Health Suboffice = Sub Department of Health of Administrative City (abbreviated to optimize scheme understandng)**

**Data collection concept**

Current health data collection is using a conventional gradual system from the smallest administrative area into a larger one. It starts from the level of Kelurahan (subdistrict), Kecamatan (district), Kotamadya (Administrative City), and at the province level (Jakarta City) as seen in Figure 3. Pusat Kesehatan Masyarakat (puskesmas) is a community health center that responsible both for health service delivery and health data collection in subdistrict and district area. Data then will be pooled at the administrative city level by Sub Department of Health of Administrative City to be validated. Those aggregate data are then sent to Jakarta Department of Health (Health department) to be analyzed further. The principal objective of this gradual system is to ensure data validity, but in recent years it has been considered to lengthen the data collection system. Otherwise, its validity is still problematic due to manual data recapitulation using paper-based reports or using offline and online Microsoft Excel in some districts.

In the first phase (2021), we planned to improve the system by giving responsibility to the Sub Department of Health of Administrative City for direct data input into a data warehouse. This period also let an introduction and adapted new data collection schemes by using a single entry for the first time. The single entry defines as only one source for specific data of which increases its reliability. Although it might be questionable as no triangulation was made, yet giving responsibility should increase the validity of the data reported. For example that malnutrition data might come from the nutrition section program, child health section program, and on many occasions both data are not in accordance.

The next phase (2022-2023) is a more open data phase that the responsibility for data input is given to sub-district and district health office. Data are then simultaneously validated and verified by the Sub Department of Health of Administrative City and Jakarta Department of Health but did not delay the data flow.

**Software and system**

The basic platform use for this system is District Health Information System 2 (DHIS-2), an open-source platform that has been used worldwide. The main purpose of using this software is data presented in the form of routine statistic data and aggregate data. DHIS-2 also had been previously used in Ministry of Health Indonesia so that any issue such as data security and integration between national and provincial government can be minimized.

Data input is according to organization units. It defines as a person in charge (PIC) for each subdistrict and district that has access to the main website. They should regularly do data imputation to the predetermined schedule. Besides data input, PIC did not have any access to other data.

**Workshop**

Workshops for trainees consist of two different targets: 1) data manager (Jakarta Department of Health and Sub Department of Health of Administrative City) and 2) data enumerator (health program person in charge). The first workshop begins with building metadata and deciding a list of users. Metadata is a definition and code for each element and variable in the DHIS-2 system. Workshop participants are program data managers so that given access to change metadata in the future, including data input, verification, validation, and also analysis. The second workshop aimed to train about data visualization and presentation. Participants were also trained to assess the quality of data.

As an evaluation, around half of participants could not tune in to the workshop materials, due to 1) online workshop (for some participants), 2) unfamiliarity to technology, and 3) difficulty to understand technical materials due to some not user-friendly dashboard according to participants.

Data quality was assessed using the principle of 4C: current, correct, consistent, and complete. Current defined as actual data by routinely update health program data. In this new system, the Sub Department of Health of Administrative City had to be more concentrated on monitoring data inputation, thus following the principle of consistent. Correct defines to ensure minimal error during data input using system validation for each variable (deciding which nominal or decimal data type, and range set for each variable). The main purpose is to reduce human error during data imputation. Data were then verified and any deviated data should be justified. Complete means that each form should be filled completely before being submitted. Evaluation

An in-depth interview is conducted with data enumerators, data monitoring (sudinkes), and program evaluator (dinkes). The major advantage of this data warehouse is the simplicity and convenience to collect a wide data from a different source and presenting it faster than using the conventional system. Less data contradiction was also found between health programs with intersecting data. We could provide a single data on an indicator that was approved by all organizational units. As long as data were collected regularly, the Jakarta Department of Health might update all relevant public health data for public consumption, as never before.

Another advantage of using DHIS-2 is that no data lost during network disconnection due to the offline data saving ability of the system. Data then be synchronized when the network available. This is crucial as internet connection in the city of developing countries, such as Jakarta, might be problematic. This would also reduce the risk of re-inputation due to connection loss, a situation that often happened.

During this transition phase, a double-work is made as data should be input to both the DHIS-2 system by Jakarta and the National Ministry of Health system, but an integration process is ongoing, and hopefully that in 202 single data entry can be established.

## Discussion

### Sustainability

The key of successful data management system is to maintain its sustainability. In previous experience at West Africa countries, a collaboration with international investigators and also school of medicine and public health are an important point of leverage to facilitate long-term retention.[4] Those collaboration might transform the quality of health data into a healthcare services delivery improvement in host-country according to Heidelberg-Nouna institutional collaboration past experience.[5] An intense training in bioinformatics also offer an advantage to maintain sustainability of data collection as this might be an obstacle found in our development of Jakarta data management system.[4] Another key of sustainability are an agreement on system vision and goals followed by innovative tools to measure this activity.[6] The goals should be delivers in clarity and concordance, thus boost the implementation of data collection system.[7]

### Other Countries

By implementing data collection management system, West Africa had been able to control malaria through improving quality of data for healthcare services. Those systems bring benefits by reducing data error rate, encourage more specific data based on regional, and help health professionals to combat disease, specifically Malaria at that study.[5]

Problems found in previous study are quite similar to our programs. There is a risk of hesitancy to complete transition to computer-based system and a parallel use of paper-based report system. Stakeholders engagement also crucial but sometimes problematic in real situation application.[6] Sometimes, attractiveness to participant are related inversely to the utilization in public health. Public health policy remain the preferred alternatives to improve data collection and maximize health benefits.[8] A report from India stated that developing data management system face some challenges: maintaining record-keeping system, standardizing the data collected, assuring data completeness and availability, and do routine backup on this data to prevent misuse public health data.[9]

A strong system of data collection might place public health as a strategic area in political point of view. A case study in Canada reported four main threat to health care sustainability: downgrading the status of public health within governments and health authorities, eroding the independence of medical officer on matters of public health concerns, limiting public health scope by combining it with primary and community care, and decreasing funding for public health. Thus, besides a system, the availability of funding, human resources, and structural requirement is crucial for public health system, especially on data management.[10]

## Acknowledgment

We would like to thank Mr. Ras Vagel and the team from the University of Oslo for assistance during this system development and made all the works follow the timeline. We also thank to Ministry of Health and Head of Jakarta Department of Health for the support during the development of this data warehouse system. This collaboration will continues until the system is fully developed.

## Future direction

There were also room for improvement in the development of this data warehouse:

1. Adapting to input data using computer-based system

2. Adapting to input data routinely according to predetermined schedule

3. Providing person in charge (PIC) to help enumerators when facing problems during data input